\documentclass[lettersize,conference]{IEEEtran}
\usepackage{amsmath,amsfonts,amssymb, amsthm}
\newtheorem{theorem}{Theorem}
\usepackage{algorithmic}
\usepackage{algorithm}
\usepackage{array}
\usepackage[caption=false,font=scriptsize,labelfont=sf,textfont=sf]{subfig}
\usepackage{textcomp}
\usepackage{stfloats}
\usepackage{url}
\usepackage{verbatim}
\usepackage{graphicx}
\usepackage{cite}
\usepackage{multirow}
\usepackage{tabularx}
\usepackage{verbatim}
\usepackage{xcolor}
\usepackage{url}

\begin{document}

\bstctlcite{IEEEexample:BSTcontrol}

\title{SampleHST: Efficient On-the-Fly Selection of Distributed Traces}

\author{
    \IEEEauthorblockN{Alim Ul Gias\IEEEauthorrefmark{1}, Yicheng Gao\IEEEauthorrefmark{2}, Matthew Sheldon\IEEEauthorrefmark{2}, Jos\'e A. Perusqu\'ia\IEEEauthorrefmark{3}, Owen O'Brien\IEEEauthorrefmark{4}, Giuliano Casale\IEEEauthorrefmark{2}}
    \IEEEauthorblockA{\IEEEauthorrefmark{1}University of Westminster, Email: a.gias@westminster.ac.uk}
    \IEEEauthorblockA{\IEEEauthorrefmark{2}Imperial College London, Email: \{y.gao20, matthew.sheldon20, g.casale\}@imperial.ac.uk}
    \IEEEauthorblockA{\IEEEauthorrefmark{3}Universidad Nacional Aut\'onoma de M\'exico, Email: jose.perusquia@sigma.iimas.unam.mx}
    \IEEEauthorblockA{\IEEEauthorrefmark{4}Huawei Technologies (Ireland) Co., Ltd, Email: owen.obrien@huawei.com}
}

\markboth{Journal of \LaTeX\ Class Files,~Vol.~14, No.~8, August~2021}%
{Shell \MakeLowercase{\textit{et al.}}: A Sample Article Using IEEEtran.cls for IEEE Journals}

\maketitle

\begin{abstract}

Since only a small number of traces generated from distributed tracing helps in troubleshooting, its storage requirement can be significantly reduced by biasing the selection towards anomalous traces. To aid in this scenario, we propose SampleHST, a novel approach to sample on-the-fly from a stream of traces in an unsupervised manner. SampleHST adjusts the storage quota of normal and anomalous traces depending on the size of its budget. Initially, it utilizes a forest of Half Space Trees (HSTs) for trace scoring. This is based on the distribution of the mass scores across the trees, which characterizes the probability of observing different traces. The mass distribution from HSTs is subsequently used to cluster the traces online leveraging a variant of the mean-shift algorithm. This trace-cluster association eventually drives the sampling decision. We have compared the performance of SampleHST with a recently suggested method using data from a cloud data center and demonstrated that SampleHST improves sampling performance up to by 9.5$\times$.  

\end{abstract}


\section{Introduction}
\label{sec:intro}

\IEEEPARstart{D}{istributed} tracing is tailored primarily to monitoring and profiling applications built with the microservice-based architecture \cite{richardson2018microservices}. In a microservice ecosystem, with the increase of services, the volume of the trace data, used for observability of application performance and reliability, increases significantly \cite{guo2020graph}. In a typical production setup, each server, hosting hundreds of microservices, generates several tens of gigabytes of trace data every day. Considering all the servers, the total daily generated data are in the order of several terabytes. Nevertheless, most of the traces do not report on application anomalies and thus there is little value in storing them all. The fraction that can be retained is constrained by a storage budget \cite{las2018weighted} and the problem we study is how to select the most interesting traces to help monitoring and diagnostics of microservices runtime behavior. This entails sampling a mix of traces that characterizes the overall user behavior but at the same time retaining a high relative ratio of anomalous traces. 

To accommodate the storage budget, we need to deploy a sampling strategy. It is a common industry practice to use uniform sampling~\cite{las2018weighted}, which is also referred as \emph{head-based} sampling. Under this strategy, the sampling decision is taken once the request for a service is received, leading to a lower hit rate of anomalous traces. To address this issue, it is increasingly preferred to use a \emph{tail-based} sampling strategy ~\cite{parker2020distributed}, which can improve the selection accuracy as it takes the sampling decision after the response is served, \emph{i.e.}, when the entire trace for the service call chain is available. This allows to reason on the information contained in the trace itself upon deciding whether to store it or not.

Ideally, a tail-based sampling strategy should be online and without any batch processing. This means that we must decide either to save or discard a trace on-the-fly rather than storing it temporarily for batch processing. Recently, researchers have proposed different tail-based sampling strategies based on unsupervised learning \cite{las2018weighted, las2019sifter, huang2021sieve}. However, existing research faces multiple challenges such as difficulties in performing clustering due to high dimensionality of data,  requirements of batch processing, low amplitude scores for anomalous traces, and no explicit consideration of the budget size. To address all these shortcomings, we propose a novel method, \textit{SampleHST}. On the one hand, SampleHST focuses on sampling only anomalous traces when the storage budget is comparatively lower than the fraction of expected anomalies. On the other hand, when the budget is higher, SampleHST samples both the normal and anomalous traces, with a bias towards anomalous ones. Such a bias is fair because it increases the representation of the anomalous traces, which are rare compared to normal ones, among the sampled traces. In other words, the bias allows  representative sampling \cite{las2018weighted, las2019sifter}.

SampleHST leverages a Bag-of-Words (BoW) model \cite{zhang2010understanding} as a count-based representation for each trace. By taking this representation as an input, we can generate a distribution of the mass values obtained from a forest of a tree-based classifier, namely Half Space Trees (HSTs) \cite{tan2011fast}. This distribution is then used to perform an online clustering of the traces based on an algorithm we have developed which is part of the mean-shift clustering algorithm family \cite{baruah2012evolving}. Once the clustering is complete, we decide to sample the trace based on its cluster association, \emph{i.e.}, a trace is more likely to be sampled if it is associated with a cluster with low mass values as such clusters represent rarely observed traces.

We evaluate the performance of SampleHST, using data provided by a commercial cloud service operator and comparing the results with a recently proposed approach for point anomalies developed in \cite{las2018weighted}. For this production dataset, we see that SampleHST yields 2.3$\times$ to 9.5$\times$ better sampling performance in terms of precision, recall and F1-Score than prior work. When we consider representative sampling in a high budget scenario, we see SampleHST is 1.6$\times$ fairer with respect to the Jain fairness index~\cite{jain1984quantitative}. In summary, the key contributions are:

\begin{itemize}

    \item A novel approach to sample distributed traces by forming clusters using the mass distribution of the traces obtained from Half Space Trees.
    
    \item An online clustering method, generalizing the mean shift algorithm \cite{fukunaga1975estimation}, that considers non-spherical cluster shapes such as hyper-cubes and hyper-rectangles.
    
    \item Experiments using real-world data to compare the sampling performance of SampleHST with a recent tail-based sampling approach \cite{las2018weighted}.
\end{itemize}

The rest of the paper is organized as follows. Section \ref{sec:background} presents the related work and motivation for developing SampleHST. Section \ref{sec:AD_trace} demonstrates how to model traces and detect anomalies. Section \ref{sec:fairhst_method} discusses how to transform anomaly detection processes to a sampling method. Section \ref{sec:cluster} and \ref{sec:fairhst:controller} present the SampleHST clustering and sampling algorithms respectively. Section \ref{sec:evaluation_point_anomalies} evaluates the sampling performance. Section \ref{sec:conc_fairhst} concludes the paper. Proofs are given in the Appendix.

\section{Background}
\label{sec:background}

\subsection{Related Work}
\label{sec:related_work_fairhst}

The first step of designing a sampler is to differentiate the anomalous traces from the normal ones. There have been many works on anomaly detection for microservices using their generated traces. The authors in \cite{wang2020workflow,zuo2020intelligent} learn from the patterns of call trees and request execution respectively to detect anomalies. Some studies \cite{nedelkoski2019anomaly2,nedelkoski2019anomaly,bogatinovski2020self} also consider deep learning based methods focusing on different aspects, \emph{e.g.,} response times and causal relationships. However, these works do not consider our sampling scenario, \emph{i.e.}, they only focus on anomaly detection but not on transforming the anomaly detection result to a sampling decision. 

To the best of our knowledge, there are only a few research papers focusing on sampling anomalous traces generated by microservices. In \cite{las2018weighted}, the authors propose a sampler based on a hierarchical clustering method PERCH \cite{kobren2017hierarchical}. Authors demonstrate that their method can achieve representative sampling, meaning equal share for both normal and anomalous traces. Such clustering methods can incur the curse of the data dimensionality \cite{zimek2012survey} and they often require batch processing, which is not always supported under low latency requirements.

Sifter \cite{las2019sifter} avoids batch processing by taking sampling decisions trace-by-trace. It generates a sampling probability by utilizing the loss of training a neural network for a particular trace. A potential issue with loss-based methods is that anomalous traces may still have small probabilities overall, closer to 0 than to 1, allowing several anomalous traces to go unsampled. This problem is studied in recently proposed sampler, Sieve \cite{huang2021sieve}, which uses a threshold to first separate the anomalous traces and then amplify the sampling probability. This still leaves an open challenge regarding the optimal and automated choice of threshold.

\subsection{Sampling performance}

As a classification problem, it may be natural to study trace sampling performance in terms of F1-Score, as this strikes a balance between Precision and Recall. We however observe that this is not always an ideal performance criterion in the presence of budget constraints. For example, an abundant storage budget with few constraints is more appropriate to consider Recall, while a heavily constrained storage budget expects more from achieving high Precision.
Summarizing, we set the following overall performance evaluation principles for trace sampling methods:
\begin{itemize}
    \item For infrequent anomalous traces, where the prevalence of anomalies is less than the storage budget, the primary evaluation metric should be the Recall.
    \item  For low storage budgets, where the prevalence of anomalies is greater than the storage budget, the primary evaluation metric should be the Precision.  
    \item When sampling $N$ traces from a collection of traces containing $N$ anomalies, the primary evaluation metric should be the F1-Score. 
\end{itemize}

\subsection{Comparing State-of-the-Art Anomaly Detection Methods}
\label{subsec:compare_ad_methods}
Since anomaly detection is a key step for a sampling process, we here illustrate why off-the-shelf anomaly detection methods are not fit for purpose. We consider the following popular techniques: 1) local density estimate: \textit{K-Nearest Neighbor} (KNN) and \textit{Local Outlier Factor} (LOF), 2) tree-based classification: \textit{Isolation Forest} and \textit{Half Space Trees} (HST) \cite{tan2011fast}, 3) boosting: \textit{Lightweight Online Detection of Anomalies} (LODA) \cite{pevny2016loda}, and 4) neural network: \textit{Deep Belief Net and One Class Support Vector Machine} (DBN+OCSVM)~\cite{erfani2016high}. A notable advantage of using the tree-based methods is that they can work on one trace at a time, while the other methods, off-the-shelf, require batching. 

To evaluate the performance of the above methods, we consider a production dataset from a cloud data center consisting of trace data spanning a week over a set of 14 microservices. As the trace is unlabelled, we identify $\sim5\%$ point anomalies using the popular offline DBSCAN clustering algorithm, and evaluate the ability of the listed methods to obtain similar results. DBSCAN, being resource intensive, is not feasible in an online scenario such as distributed trace sampling, but is considered as a generally reliable technique in industry \cite{netflix2015tracking}. We use Matlab's native implementation of DBSCAN with $\epsilon=2.5$ and $minpts=5$, where $\epsilon$ indicates the size of the local neighborhood of the data points and $minpts$ indicates the minimum number of points per cluster. Once the traces are clustered, we regard the smallest clusters as anomalies, accounting for $\sim5\%$ of the total traces. 

\begin{table}[t]
\footnotesize
\centering
\caption{Results of different anomaly detection methods on the production dataset}
\label{tbl:result-nce-march02-include15}
\begin{tabular}{|l|c|c|c|c|c|c|}
\hline
  &
  \textbf{\begin{tabular}[c]{@{}c@{}}Isolation\\ Forest\end{tabular}} &
  \textbf{KNN} &
  \textbf{LOF} &
  \textbf{LODA} &
  \textbf{\begin{tabular}[c]{@{}c@{}}DBN +\\ OCSVM\end{tabular}} &
  \textbf{\begin{tabular}[c]{@{}c@{}}HST\end{tabular}} \\ \hline
   Precision & 0.73 & 0.77 & 0.73 & 0.62 & 0.47 & 0.94 \\ \hline
   Recall    & 0.72 & 0.72 & 0.72 & 0.60 & 0.97 & 0.70 \\ \hline
   \textbf{F1-Score}   & \textbf{0.73} & \textbf{0.74} & \textbf{0.73} & \textbf{0.61} & \textbf{0.64} & \textbf{0.80} \\ \hline
\end{tabular}
\end{table}

The results of the experiment are presented in Table \ref{tbl:result-nce-march02-include15}. The dataset contains traces from six consecutive days with 77577 traces. For all the batch methods, we keep a similar batch size of 2000 traces. We see that HST is the best method with respect to F1-Score. This motivates further investigation in HST methods to address the problem under study. In addition, HST has other benefits from the perspective of a streaming platform. Due to the way HSTs are designed, for a particular trace, we only need to update a single mass~value~\cite{ting13massest}  per tree. To determine whether a trace is normal or anomalous, the mean mass value ($m$) of the HSTs, for that particular trace, is compared against a threshold. An HST only needs to query its already stored mass values, resulting in a very low computational footprint in the order of less than a millisecond per trace. This will reduce the time taken during the training, where we can only use the computing resource to update the mass values of the node. Due to all these benefits, the rest of the paper focuses on HST as a baseline classifier.

\section{Half Space Trees for Anomaly Detection}
\label{sec:AD_trace}

Half Space Trees (HST) \cite{tan2011fast} are an ensemble of decision trees. The structure of the decision trees is a simple Binary Tree. Each HST has a depth $d$, and the corresponding binary tree will have $2^{d+1}-1$ nodes. Each tree stores split points for a random subset of dimensions, and possibly multiple splits per dimension, together with a count of how many points are within the subspace defined by a path (a metric called {\em mass}). Mass is simply defined as a count of data points, thus it is easier to calculate than density measures used in other methods, \emph{e.g.}, which require likelihood estimation. Normally, an ensemble of $t$ Binary Trees is used, with identical depth $h$, which are independently trained on a data window $w$.

HSTs are particularly suitable for streaming data as its core processes - building the tree data structure and characterizing the data points using the mass values - are both lightweight \cite{tan2011fast}. In this study, we assume that such data points will be available of continuously arriving streams of spans generated in a cloud data center from a heterogeneous collection of microservices. A span is an immutable data structure that supplies the value of a collection of categorical and continuous variables at a particular point in time. The spans contain a \textit{traceId}, based on which they can be grouped to form traces. We propose to abstract each trace as a document where the span properties are considered as words or terms. The document is subsequently converted to a bag of words \cite{zhang2010understanding}.

During the conversion, the span properties that are not relevant to performance and reliability analysis are ignored. We restrict our attention to discrete fields, some of which, \emph{e.g.}, HTTP code, can be categorical \emph{i.e.}, they have a fixed number of possible values. In addition, we do not explicitly address latency anomalies as they are often best studies with with anomaly detection based on continuous response time distribution estimators, which can be already done with specialized methods in the literature\cite{li2022longtale,wu2020microrca,meng2020localizing}. Alternatively, latencies can be discretized and considered as one of the features considered by our method. We represent each trace using a count vector $x=(x_1,\ldots,x_d,\ldots,x_D)$, where $D$ is the number of different terms that have been seen across all the traces. For example, the HTTP code 200 is one term and a specific URL could be another one. Each dimension $x_d\geq 0$ is an integer value counting how many time a particular term appears in a trace. The resulting count data assures knowledge of the dimension $D$ and the mappings of dimensions to terms. In a production implementation, such knowledge can be acquired from an initial monitoring period and periodically updated.

In production data, sparsity is frequently observed. Once the categorical properties of the spans are vectorized as count data, there are relatively few types of traces that occur repeatedly, thus the HST mass could accumulate within a small set of terminal nodes. This is confirmed from our production data where we observe that only $0.003\%$ of the trace count vectors are unique. We thus focus on a variant of HST known as HS*-Trees (HS*T) \cite{ting13massest}, which aims to deal with the sparsity in the tree structure. In HS*T, nodes that have fewer than \textit{SizeLimit} samples are not further expanded during the training phase. This reduces memory consumption and also the time to traverse the trees. Thus, we have opted for HS*T as our chosen HST variant. In the rest of the study, we use the term HST and HS*T interchangeably.

We incorporated two further modifications to HS*T. Firstly, we opted for depth-dependent split dimension. This means that when splitting a node, instead of using the normal procedure of picking a dimension at random, we require all nodes at the same depth level to use the same split dimension, which largely reduces memory usage since a single dimension is stored at each level. Secondly, as suggested in \cite{tan2011fast}, we opted for a $[0,1]$ workspace. This means that, the maximum and minimum values of the features are assumed by the HS*T to be 1.0 and 0.0, rather than in the min-max range observed in the data. This can simply be achieved with min-max scaling. However, an issue with such count data scaling is that outliers can often cluster the normal values at one end of the range, making the prediction particularly difficult for tree based methods since they rely on randomized partitioning of the input space, \emph{i.e.}, random split points will be chosen in the segment [0,1] to branch the tree along a dimension. Therefore, if the points are all clustered in a small portion of the range [0,1] the HS*T will struggle to separate the samples along that dimension. To address this, we apply the following transformation in place of the min-max scaling

\begin{equation}
\label{eq:transform}
f(x) = \frac{1}{1+g(x)}.
\end{equation}

\begin{figure}[t]
    \centering
     \subfloat[][Min-Max Scaling]{ \includegraphics[width=0.5\linewidth]{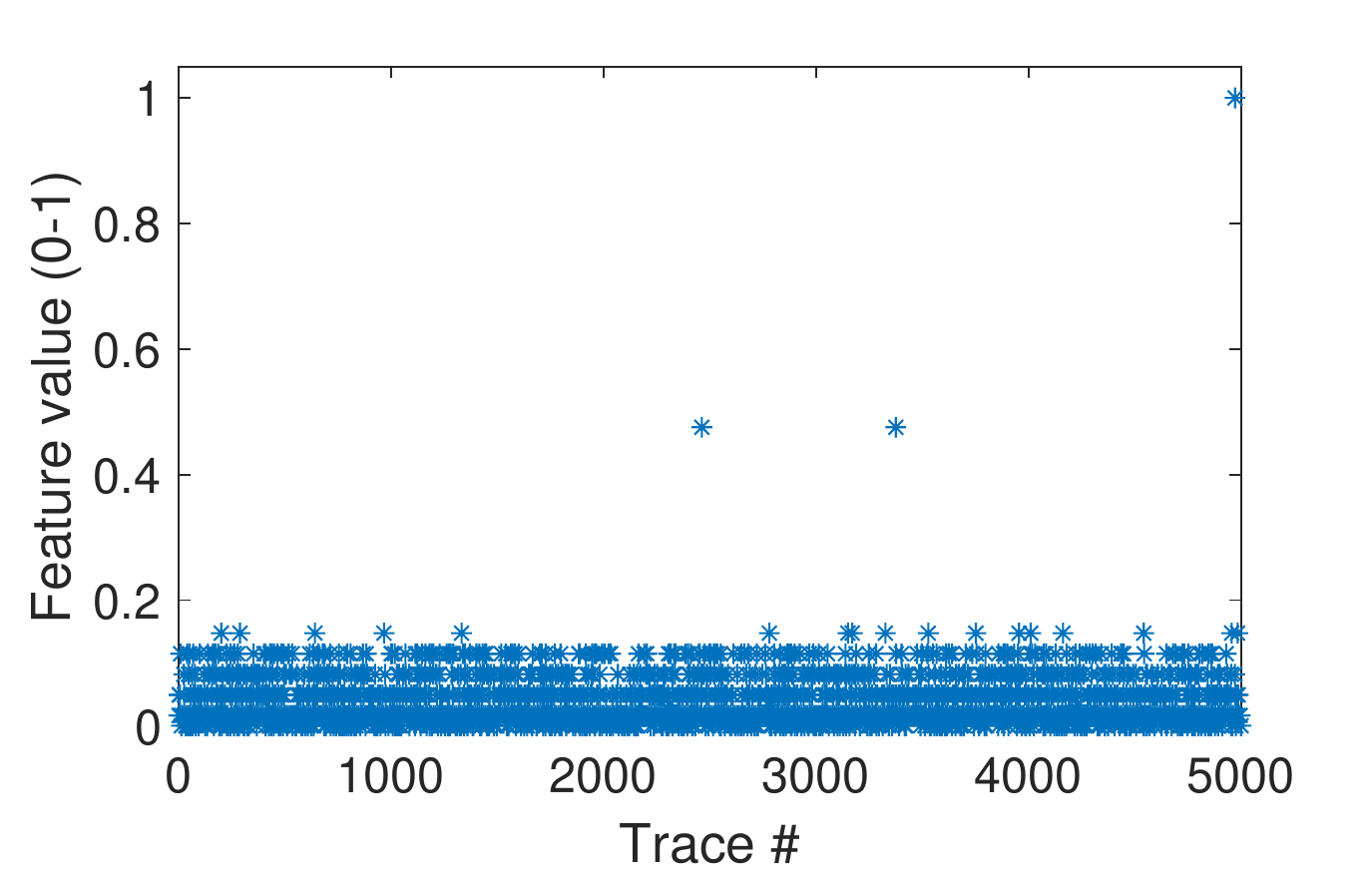}\label{fig:feature-vector}}
     \subfloat[][Transformation function $f$]{\includegraphics[width=0.5\linewidth]{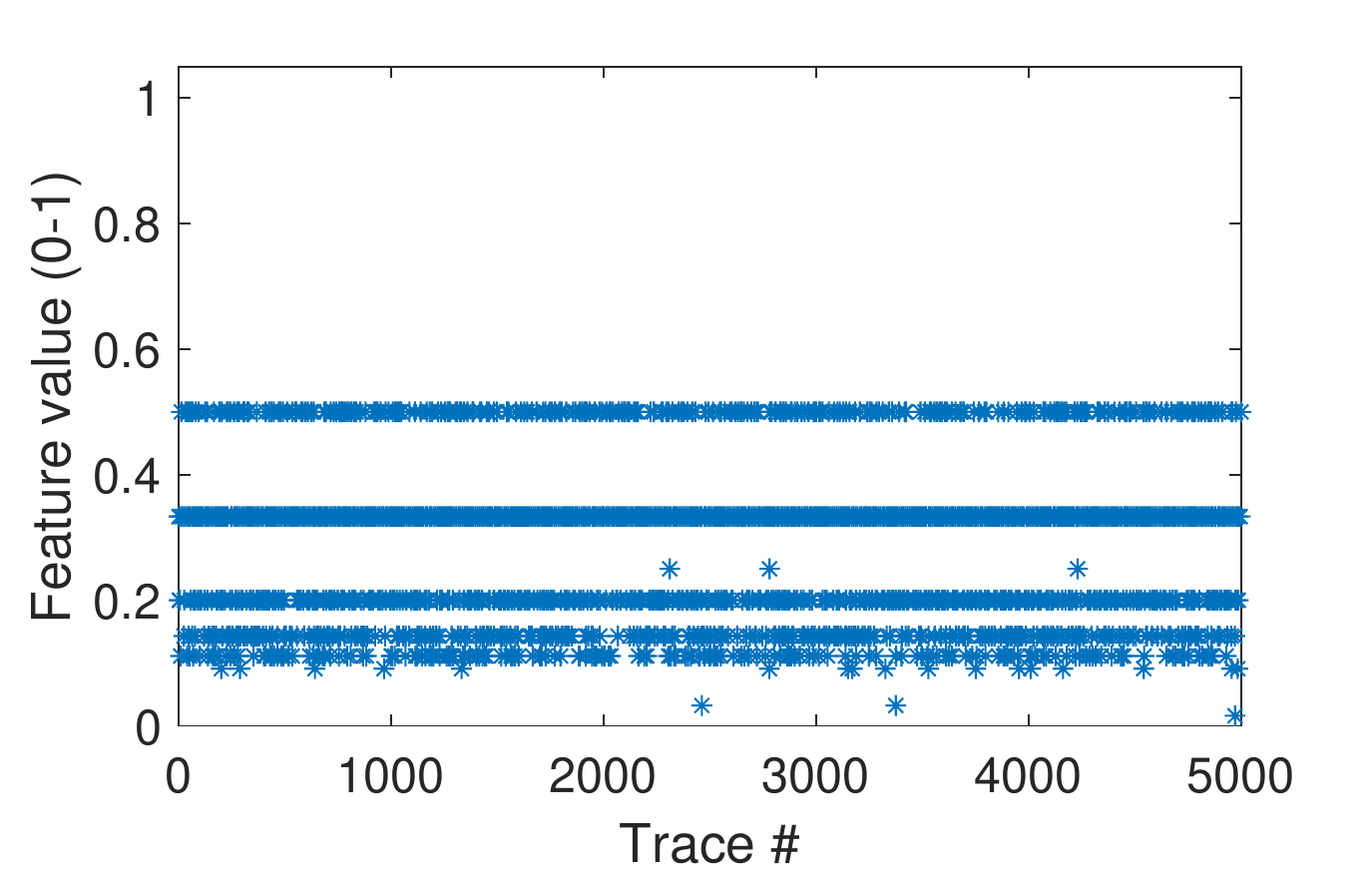}\label{fig:feature-vector-transform}}
    \caption{Comparing the scaled value of HTTP 200 code counts with min-max scaling and the transformation function $f$}
    \label{fig:feature-vector-transform-compare}
\end{figure}

that allows us to control the stratification of the count data. We have found it sufficient to use $g(x) = x$ but we could also define, for example, $g(x) = \log(x)$ considering large values of $x$. Using \eqref{eq:transform}, the large outliers will be squeezed near 0, therefore not suppressing the ability to resolve the normal values That are critical to HST training. We illustrate the impact of this transformation in Fig.~\ref{fig:feature-vector-transform-compare} using 5000 randomly chosen traces, where we scaled the feature corresponding to the frequency of HTTP code 200 in the trace.

As before, we used the production data from Section \ref{subsec:compare_ad_methods} to test these modifications. We consider each day as a window and use the first day to build the trees. We observe that the F1-Score improves from 0.8 to 0.97. This indicates that the changes aid in anomaly detection from the trace streams. 

\section{Mass-based Clustering for Sampling}
\label{sec:fairhst_method}

Although HSTs can help in classifying the anomalous traces, in reality we need to utilize this classification output
in a sampling process. This process is complex because of the trade-off between sampling normal and anomalous traces. While sampling, the proportion of the storage budget and the expected percentage of anomalies should be taken into account. If the budget is lower than the anomaly percentage, the focus should be on sampling mostly the anomalous traces. The normal traces should gain more attention only when the budget is higher than the anomaly percentage. In addition, while sampling the anomalous traces, the target should be representative sampling from that group of traces \emph{i.e.} sampling from different ``groups'' of traces fairly. 

To achieve this, we propose to cluster the traces and decide whether to sample a trace or not based on its cluster association. However, when clustering in a high-dimensional space it is harder to achieve accurate density estimation \cite{fukunaga2013introduction}, in addition to incurring a higher computational cost. This is expected in the normal behavior of our system, as our production data contains hundreds of features. Therefore, we propose a new approach considering the distribution of mass across the trees in the HS*T forest and selecting a mean mass score $m$ and a low percentile of the mass score $p$. Low percentiles are expected to significantly differ from the mean when there is at least a subset of trees in the forest that identifies the trace as an anomaly. We refer to this method as \emph{SampleHST} as we are using the the mass distribution of HST to perform sampling. 

\begin{figure}[t]
    \centering
    \includegraphics[width=0.66\linewidth]{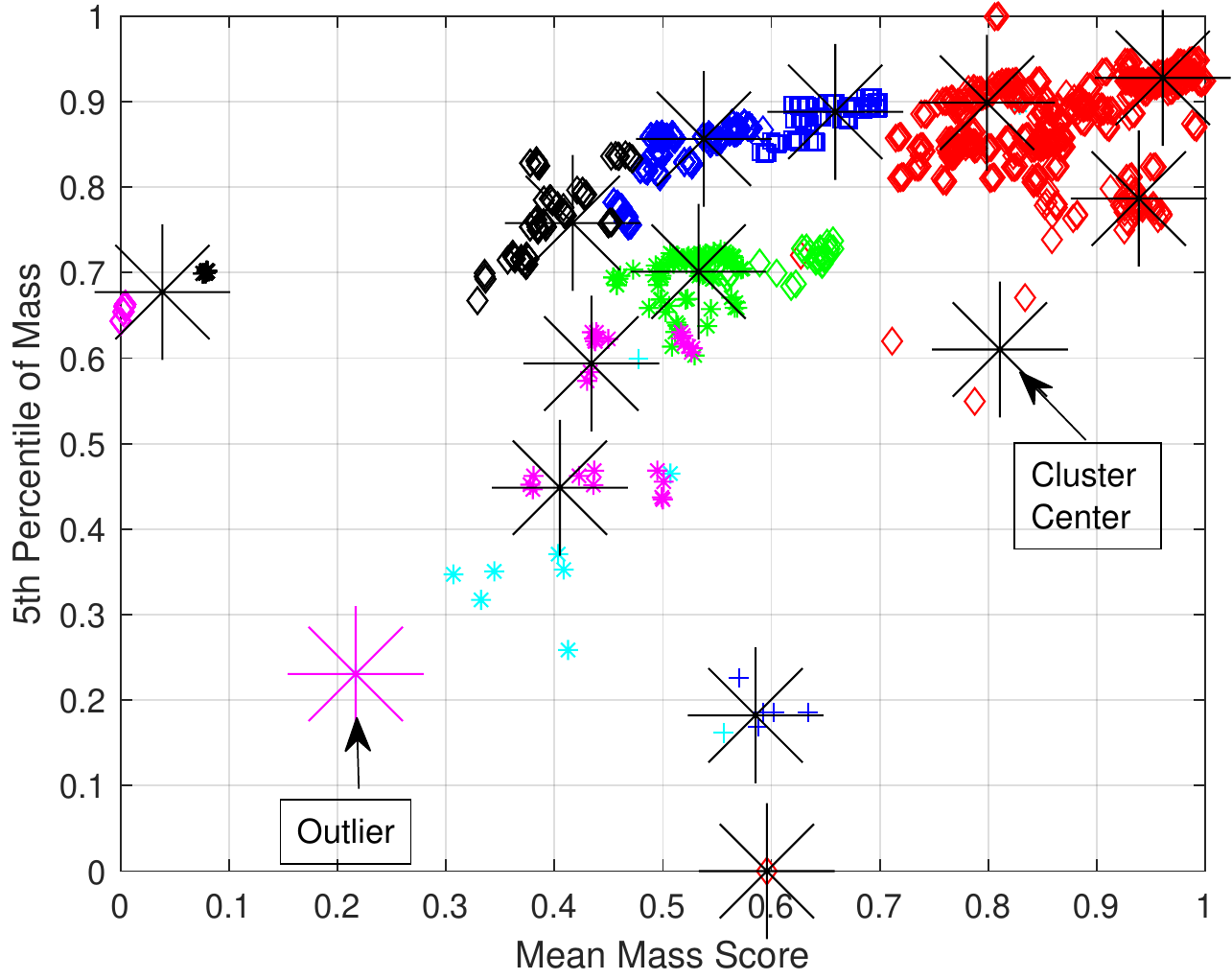}
    \caption[The production trace plotted using the mass-based properties]{The production trace plotted using the mass-based properties. The colors and marker shapes indicate the DBSCAN original clusters. The cluster centers are estimated with a baseline online clustering method.}
    \label{fig:hst_cluster}
\end{figure}

Since we want to use a low percentile ($p$) value along with the mean ($m$), we represent each trace with a unique pair $(m,p)$ that will be used for clustering. The projection of the production traces from Section \ref{subsec:compare_ad_methods} in this 2-dimensional space is shown in Figure \ref{fig:hst_cluster}. The figure shows in different colors the clusters obtained by DBSCAN. It is clearly seen that the mass-based properties cluster the traces in distinct groups and the cluster centers are also appropriately detected using a baseline streaming clustering method \cite{hyde2017fully}. Another potential benefit of using the low percentile value is a better separation of trace groups. As seen from Fig. \ref{fig:hst_cluster}, ignoring the percentile value will result in multiple trace groups being merged together, eventually affecting the sampling performance. 

This mass-based clustering is at the core of our sampling approach. Once a trace is formed with its spans, to take a sampling decision, it is moved through two key components:
\begin{itemize}
    \item \textbf{SampleHST Clustering}: Cluster the trace based on the its mass based properties.
    \item \textbf{SampleHST Controller}: Makes the sampling decision based on budget and trace-cluster association.
    
\end{itemize}
We discuss these components in details in the next sections.

\section{SampleHST Clustering}
\label{sec:cluster}

\textit{SampleHST Clustering} is primarily based on the underlying theory of \emph{mean-shift analysis} \cite{baruah2012evolving} and the CEDAS algorithm \cite{hyde2017fully} yielding a data-driven online approach that generalizes the hyper-sphere cluster shape commonly assumed in the literature to hyper-rectangles and hyper-cubes. Broadly speaking, our method receives the mass score of a trace in the form of a pair $(m,p)$, which is generated using the HST mass distribution. Subsequently, the method aims to find the association of the new trace with an existing cluster, if the association condition is not met a new cluster is created and a signal is send. Furthermore, the method is able to remove clusters that have not received a new trace for a pre-defined period of time modulated by the \emph{decay} and the \emph{life (energy)} parameters and also merge clusters together whenever an overlapping occurs. These steps can be broadly grouped into two sets of tasks: \emph{trace association} and \emph{cluster management}. We now discuss the key aspects of these tasks.

\subsection{Trace Association}

\subsubsection{Cluster Shape}

A common assumption for online clustering algorithms for data streams is that the cluster shape is a hyper-sphere \cite{baruah2012evolving, hyde2017fully}. In our case, the problem with such shapes is that they can lead to inaccurate partitioning of the traces because the normalized values of the unique pair $(m,p)$ belong to the unit hyper-cube. To address this issue we consider instead an arithmetic average kernel whose support is a hyper-rectangle \cite{langrene2019fast}. Assuming $d$ data dimensions, the kernel considered is presented in \eqref{eq:additive_kernel} for which we further show in Theorem~\ref{Theorem1} that the mean-shift property is achieved if the clustering bandwidth \cite{wand1994kernel} is equal in all dimensions.

\begin{theorem}
\label{Theorem1}
For the additive kernel defined as 
\begin{equation}
\label{eq:additive_kernel}
  K_d(u_1, \ldots u_d) =
    \begin{cases}
      \frac{3}{d2^{d+1}}\sum\limits_{k=1}^d \left( 1-u_k^2\right) & \text{if $|u_{k}| < 1, \forall k$}\\
      0 & \text{otherwise},
    \end{cases}       
\end{equation}
the mean-shift algorithm at each iteration shifts each sample with a value equal to the local mean if the support is given by a hyper-cube.

\end{theorem}

The proof is given in the Appendix.

\subsubsection{Cluster Assignation}
\label{subsubsec:cluster_creation}

The assignment step requires a pre-defined clustering bandwidth. We define the bandwidth vector, $H=\{ h \in \mathbb{R}^d | \forall i =1,\ldots, d, \ 0 < h_i \leq 1\}$, where each value $h_i \in H$ defines the Manhattan distance from the center to the boundary of the cluster in the $i^{\text{th}}$ dimension. Now, if we define a vector of Manhattan distances between a cluster centroid and a new data point as $M=\{ m \in \mathbb{R}^d | \forall i =1,\ldots, d, 0 \leq m_i \leq 1\}$, then if $\forall i \  m_i \leq h_i$, we assign the data point to that cluster. Otherwise, a new cluster is created with that point.

\subsubsection{Centroid Update}
\label{subsubsec:centroid_update}

Appropriately updating the cluster centroid is critical since \emph{SampleHST} uses the centroid distance to decide the mapping of traces to clusters. In general it is preferable to update the centroid giving more importance to traces that are unequivocally within that cluster. This is the concept of \textit{cluster kernel region} \cite{hyde2017fully}. Given the clustering bandwidth vector $\mathbf{H}$, we can define the kernel region as the sub-space within a cluster with bandwidth $r\mathbf{H}$, where the scalar $r$ quantifies the proportion of the cluster considered as the kernel region.

\subsection{Cluster Management}

\subsubsection{Cluster Merging}
\label{subsubsec:cluster_merging}
\begin{figure}[t]
    \centering
    \includegraphics[width=\linewidth]{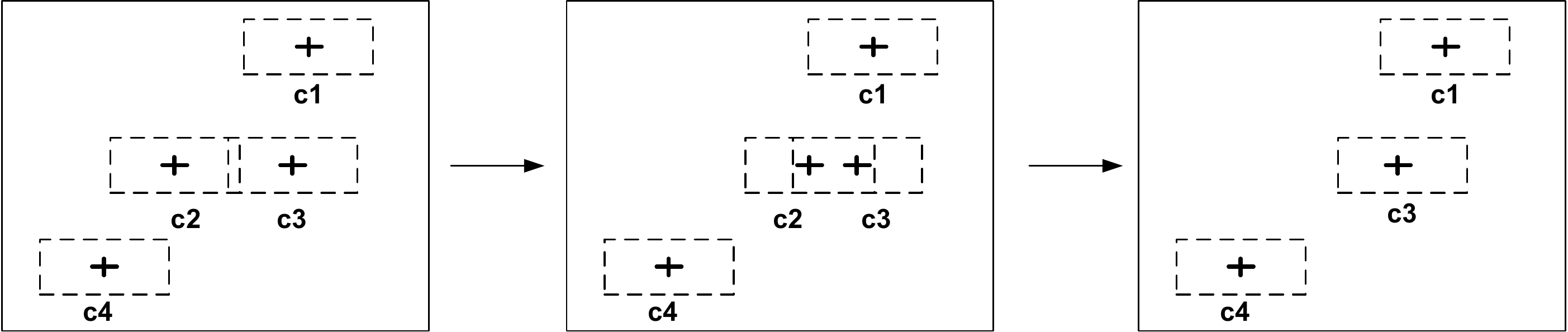}
    \caption[Demonstrating cluster merging process]{Demonstrating cluster merging process. Initially, though there is an overlap between the boundary of cluster \texttt{c2} and \texttt{c3} they are not merged. Once their centroids overlap, they are merged into single cluster \texttt{c3}}
    \label{fig:clust_merging}
\end{figure}

To address the overlaps among clusters as they are indications of possibly inaccurate clustering, we opt for the policy that merges two clusters only when the centroid of one overlaps with the boundary of the other. This policy is less drastic than merging two clusters when their boundaries overlap because one distant point cannot shift the cluster center unless the cluster has a very few samples. An illustration of this policy is presented in Fig. \ref{fig:clust_merging}.

\subsubsection{Cluster Removal}
\label{subsubsec:cluster_freshness}

We need to regularly remove the clusters whose population have remained static for a while since they are unlikely to be relevant and might affect the sampling policy. We realize this by using the \textit{decay} and \textit{life (energy)} parameters for the clusters as in \cite{hyde2017fully}. The life property is initially set to one and gradually reduced using the decay value, which is set as the average number of traces in the work cycles, defined as a sequence of consecutive periods where we received at least $1$ trace, within the sampling window.

\section{SampleHST Controller}
\label{sec:fairhst:controller}

\subsection{Overview}

The SampleHST controller ultimately decides whether to sample a trace or not by utilizing the clustering method we have presented. The controller initially calculates the number of traces ($s_w$) that need to be sampled from the next sequence of $w$ traces. We refer to this number as sampling limit and the sequence as a window. For a given budget $\tau$, the sampling limit is defined as $s_w = \tau w$. The budget is held constant, therefore the sampling limit only varies with $w$ over the runtime. The sampling process runs continuously according to Algorithm~\ref{alg:fairhst_controller}, using HST mass scores $x_m$. The algorithm expects a set of inputs that defines the size of the sampling window ($w$), the budget ($\tau$), the total number of traces to be sampled in this window ($s_w$), the relative position of the current trace in the window ($w_i^{(p)}$), the number of traces that still remain to be sampled ($s_r$), the current clusters status ($C$), the clustering bandwidth vector ($H$) and the length of the work cycle ($\beta$).

The first step in this algorithm is to make any needed adjustments to the sampling window size estimate and the sampling target. This is followed by log-transformation and min-max scaling of mass scores: $x_m^{(s)} = [log_b(x_m) - \min(x_m^{s})]/[\max(x_m) - \min(x_m)]$. The cluster centers are re-scaled only if the minimum or maximum values change along with the new mass scores in the sampling window. Once this pre-processing part is completed, the locality of the trace, represented by the cluster which in it falls, is determined by the SampleHST clustering method. If the sampling target is not reached, the final step is to make a decision based on the inclusion of the trace in a set of prioritized clusters, which we refer as the \textit{selection pool}. This step is skipped if the sampling target has already been reached.

\begin{algorithm}[t]
\caption{Sampling Process}
\label{alg:fairhst_controller}
\begin{algorithmic}[1]
\REQUIRE massScores ($x_m$), budget ($\tau$), idxPriWindow ($w_i^{(p)}$), windowSize ($w$), remainingTarget ($s_r$), windowTarget ($s_w$), clusters ($C$), bandwidth ($H$), workCycleLen ($\beta$)
\ENSURE $decision$
\IF {$w_i^{(p)} > w$}
    \STATE $\text{AdjustParameters}()$
\ENDIF
\STATE $x_m^{(\log)} = \log_b(x_m)$
\STATE $x_m^{(s)} = \text{ScaleScores}(x_m^{(\log)})$
\IF {$\text{HasMaxMinChanged}()$}
    \STATE $\text{ReScaleClusterCenters}()$
\ENDIF
\STATE $(C, x_c) \leftarrow \text{GetTraceLocality}(x_m^{(s)}, C, H, \frac{1}{\beta)}$
\STATE $R = \frac{w_i^{(p)}}{w}$
\STATE $U = \frac{s_w-s_r}{s_w}$
\IF {$s_r > 0$}
    \STATE {$decision = \text{IsTraceInSelectionPool}(C, x_c, \tau, R, U)$}
\ENDIF
\IF {$decision$}
    \STATE $s_r = s_r - 1$
\ENDIF
\end{algorithmic}
\end{algorithm}

Since we already discussed the SampleHST clustering method, we only present the other key aspects of the controller in the following sections.

\subsection{Online Score Scaling}
\label{subsubsec:min_max_scaling}
The SampleHST clustering method uses the mean and the $5^{\text{th}}$ percentile of the mass to cluster the traces ($p=0.05$). Since we are using HS*T, we use the  mass value $m[l]2^l$, where $m[l]$ is the mass of the terminal node where the trace falls into and $l$ is the depth of the corresponding tree node. To standardize the mass scores, we scale down the augmented mass using the maximum mass value possible, which is $w2^d$ where $d$ is the tree depth and $w$ is the number of observed traces.

\subsection{Sampling Decision}
\label{subsubsec:sampling_decision}

The sampling decision procedure needs to decide on-the-fly whether to sample a trace or not. If a new cluster is created by a trace the methods always sample it. For the case where the trace is associated with an existing cluster, we rely instead on generating a prioritized pool of clusters, which we refer as \textit{selection pool} and use it to take the decision. This is done in three steps, which are described as follows.

\subsubsection{Distance-based Cluster Ranking}
\label{subsubsec:distance_rank}

The first step is to rank the clusters. Two methods of ranking were considered: size of the cluster and Euclidean distance from the origin. Cluster size is an obvious method of ranking, but since SampleHST creates and deletes clusters online, smaller clusters might not always represent less frequent traces. A cluster might be smaller but all of its traces can have high mass values. This means that the traces have hit HST nodes with a high mass count which indicates that these traces are quite frequent. In addition, the most interesting and possibly smallest clusters are likely to be near the origin, which represents a low mass region in the clustering place. Therefore, we chose Euclidean distance of the centroids to the origin $(0,0)$ and if a cluster is closer to the origin, traces associated with it will be sampled first even if that cluster is not the smallest.

\subsubsection{Selection Pool}

Once the clusters are ranked, we decide how many of those will form the initial selection pool. Clusters are added according to the above ranking, starting with the one closest to the origin, until the threshold $\theta$ is reached. If two clusters are equidistant, the one created first is prioritized.

After creating the initial selection pool, we start the second phase by checking the actual value of the percentage total population in the selection pool denoted by $\hat{\theta}$. If the actual percentage is less than $\alpha\%$ of the budget, we add more clusters in the selection pool. The clusters are added depending of the magnitude $M$ of the budget ($\tau$) in comparison to $\hat{\theta}$. This is defined as $M = \left \lfloor{(\tau - \hat{\theta})/{\hat{\theta}}+\frac{1}{2}}\right \rfloor$. We then make $M$ independent attempts to add the clusters in a probabilistic manner, where in the $k^{\text{th}}$ attempt, the $k^{\text{th}}$ closest cluster to the origin, which is not yet included in the selection pool, is chosen with a probability $P^k$. Here each attempt of being successful has the same probability $P = \max(\tau, S)$, where $\tau$ is the budget and $S$ is the sampling eagerness defined as

\begin{equation}
\label{eq:eagerness}
 S = R(1-U).    
\end{equation}

This sampling eagerness is bounded between $[0,1]$ and a high value indicates to sample more. It is defined in terms of the budget utilization ($U$), which is the ratio of number of sampled traces to the sampling limit, and the relative trace position in the current window ($R$), which is the ratio of the trace index in the current window to the sampling window size.

\subsubsection{Decision Process}

After the selection pool has been decided, we sample the new trace only if it is associated with any of the clusters in the pool. If that is the case, one of two paths may be followed.
If the budget is greater than or equal to the actual percentage of population in the selection pool ($\tau \geq \hat{\theta}$), we sample the trace straightaway. Conversely, if the budget is less than the actual percentage, we follow the second path that takes a probabilistic sampling decision. This is to sample cautiously as we may have larger clusters in the selection pool containing common traces. In this path, we set the probability of sampling as

\begin{equation}
\label{eq:sampling_prob}
  P_s =
    \begin{cases}
      \dfrac{\tau}{\hat{\theta}} & \text{if $\Gamma_{c} > \Gamma_{\mu} + k\Gamma_{\sigma}$}\\
      1 & \text{otherwise}.
    \end{cases}       
\end{equation}

Here we set the probability based on the cluster size. Firstly, if the size of the cluster ($\Gamma_c$), which is associated with the current trace, is greater than the sum of mean ($\Gamma_{\mu}$) and $k$ standard deviation ($\Gamma_{\\sigma}$) of the cluster size in selection pool, we set the sampling probability to $\tau/\hat{\theta}$. This means that, if there are $N$ traces, the size of the selection pool will be $N\hat{\theta}$ and we would like to sample $N\tau$ traces from those in the selection pool. Secondly, if $\Gamma_c \leq \Gamma_{\mu} + k\Gamma_{\sigma}$, we set the sampling probability to $1$. This means if the cluster is sufficiently small, we decide to sample the corresponding trace. The value of $k$ is set using Chebyshev’s inequality \cite{feller2008introduction}, which estimates the minimum percentage ($V$) of values within $k$ standard deviation of the mean. For a given $V$, we can solve the inequality to determine the value of $k$. We notice that, this percentage $V$ is related to the ratio of $\tau/\hat{\theta}$. Because, if $\tau$ is much smaller than $\hat{\theta}$, we want to sample only if the associated cluster is smaller than the majority of the clusters. As the value of $\tau$ increases compared to $\hat{\theta}$, we can consider the larger clusters i.e., larger value of $V$. Thus, we consider $\hat{V} = \tau/\hat{\theta}$, where $\hat{V}$ is and estimate of the minimum percentage $V$, we can calculate the value of $k$ using~\eqref{eq:chebyshev_ineq}. 

\begin{equation}
\label{eq:chebyshev_ineq}
k = \sqrt{\dfrac{1}{1-\dfrac{\tau}{\hat{\theta}}}} \equiv \sqrt{\dfrac{\hat{\theta}}{\hat{\theta}-\tau}}.
\end{equation}

\section{Sampling Performance}
\label{sec:evaluation_point_anomalies}

\subsection{Experimental Setup}
\label{subsec:fairHST-results-exsetup}
To test model performance we use a dataset provided by a cloud data centre composed of 77,577 traces. Each trace contains at least one span and the following four categorical features: \textit{Service Name, URL, Process Id}, and \textit{Node Id}. The dataset includes 17 different services with six of them containing 98.59$\%$ of the spans; more than 40 different URLs with two accounting for $98.55\%$ of the spans; more than 50 different Process Id's with six containing 88.31$\%$ of the spans; and 8 different node Id's with two containing 86.8$\%$ of the spans. The traces are represented as a count vector the Bag of Words model as detailed in Section~\ref{sec:AD_trace}. Through this, we obtain 105 unique features. Ignoring timestamps, the 77,577 traces map to 308 unique traces. 

To test the SampleHST robustness, we consider 5 cases with different storage budgets. First, since we have about 5\% anomalies in our data, we include a case where the budget is 5\%. The evaluation criteria for this case is the F1-Score. We have also chosen 3 smaller budgets (0.5\%, 1\% and 2\%) where the evaluation criteria is precision. Finally, we also consider a high budget case of 10\%, where the evaluation criteria is recall. We compare the results with two other samplers: uniform random sampler, implemented following the Head-based sampler in \cite{las2019sifter}, and the PERCH-based method~\cite{las2018weighted}. 

Since sampling methods such as \cite{las2018weighted, las2019sifter} focus on representative sampling, we also compare their fairness in terms the Jain index \cite{jain1984quantitative}. The index can be calculated using \eqref{eq:jain-index} where $X_i = \frac{T_i}{O_i}$. Here, for each cluster $i$, $T_i$ is the number of traces sampled by a method and $O_i$ is the optimal number of traces that should be sampled.  This metric indicates what percentage of the groups are treated fairly. In our case, the groups are the clusters that we obtain offline from DBSCAN clustering. Note that, to calculate the index, we need to know the optimal number of traces that should be sampled.  As we know the overall distribution of the traces among the groups and sampling budget, we calculate it offline using the max-min fair allocation approach \cite{jaffe1981bottleneck}. 
\begin{equation}
\label{eq:jain-index}
\mathcal{J}(X_1, X_2, \ldots, X_n) = \frac{\left(\sum\limits_{i=1}^n X_i\right)^2}{n \sum\limits_{i=1}^n X_i^2}\quad  X_i \geq 0
\end{equation}

\begin{figure}[t]
    \centering
    \includegraphics[width=\linewidth]{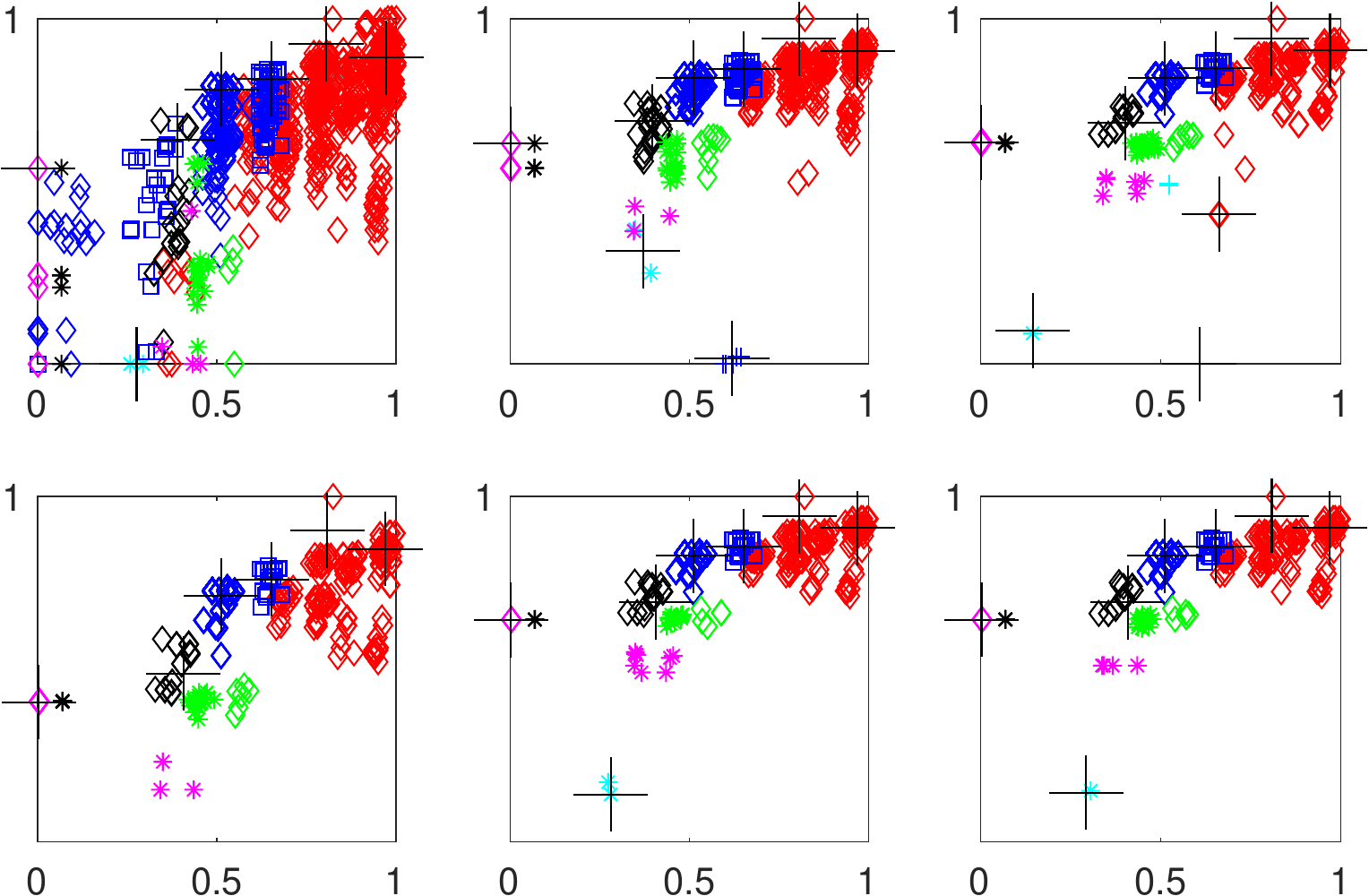}
    \caption[Illustrating the output of the SampleHST clustering algorithm]{Output of the SampleHST clustering algorithm. The X-axis and Y-axis represent mean and $5^{\text{th}}$ percentile of mass respectively. The colored symbols represent different DBSCAN labels. The {\large+} signs are the cluster centers estimated by the SampleHST clustering algorithm. The output is presented in 6 windows. As we move from left to right, we move towards the next window. }
    \label{fig:fairHST_clustering_out}
\end{figure}

\subsection{Results}
\label{subsec:fairHST-results}

\noindent {\bf SampleHST Clustering Operation.} We begin by illustrating in Fig. \ref{fig:fairHST_clustering_out} the operation of the SampleHST method. Since, this is an online clustering method, we divide the total time frame in 20 periods and show the clustering status for those periods. We immediately see that in the first window, the data points are less segregated. This is because of the online min-max scaling. In the initial period, the min-max values are not steady, which affects the data points as well. As we progress towards the end, we can see that the clusters are increasingly segregated. We also see that the number of clusters continue to change throughout these periods. The clusters around the top right corner remains stable, but the ones around the bottom left corner change their positions frequently as the top right clusters are of frequent traces whereas the bottom left ones are of the infrequent ones. The infrequent trace clusters decay quickly by not receiving traces in some work cycles. 

\noindent {\bf Comparative experiments.} We now compare the performance of SampleHST against the uniformly random and PERCH-based methods. In Table~\ref{tbl:results-compare-march02} we see that SampleHST with a bandwidth of $h=0.1$ is the best method across all budgets, with the uniform random sampler performing the worst. We also see that the PERCH-based method does not perform significantly better with respect the precision, recall and F1-Score. From the fairness perspective, the PERCH-based method scores much higher than the random sampler, but still it cannot outperform SampleHST. The results show that even though the PERCH-based method can achieve better Jain score in low budgets, it is not precise in sampling the anomalous traces as made evident by the precision score. 

As we mentioned earlier, identifying anomalous traces is difficult for clustering methods due to the high number of dimensions of the input data, as in the present case with 105 dimensions. SampleHST, on the other hand, eliminates this problem by using the mass scores, which are low dimensional.

\begin{table}[t]
\centering
\caption{Performance of different samplers with different budget}
\label{tbl:results-compare-march02}
\begin{tabular}{|cc|c|c|c|c|c|}
\hline
\multicolumn{2}{|c|}{}                                                                                               & \textbf{0.5\%} & \textbf{1\%}  & \textbf{2\%}  & \textbf{5\%}  & \textbf{10\%} \\ \hline
\multicolumn{1}{|c|}{\multirow{4}{*}{\textbf{Uniform}}}                                                 & \textbf{J}  & 0.10            & 0.10          & 0.11          & 0.13          & 0.18          \\ \cline{2-7} 
\multicolumn{1}{|c|}{}                                                                                 & \textbf{P}  & \textbf{0.05}  & \textbf{0.04} & \textbf{0.06} & 0.05          & 0.05          \\ \cline{2-7} 
\multicolumn{1}{|c|}{}                                                                                 & \textbf{R}  & 0.01           & 0.01          & 0.03          & 0.05          & \textbf{0.10}  \\ \cline{2-7} 
\multicolumn{1}{|c|}{}                                                                                 & \textbf{F1} & 0.01           & 0.01          & 0.04          & \textbf{0.05} & 0.06          \\ \hline
\multicolumn{1}{|c|}{\multirow{4}{*}{\textbf{\begin{tabular}[c]{@{}c@{}}PERCH-\\ based\end{tabular}}}} & \textbf{J}  & 0.32           & 0.24          & 0.32          & 0.47          & 0.56          \\ \cline{2-7} 
\multicolumn{1}{|c|}{}                                                                                 & \textbf{P}  & \textbf{0.41}  & \textbf{0.18} & \textbf{0.13} & 0.11          & 0.09          \\ \cline{2-7} 
\multicolumn{1}{|c|}{}                                                                                 & \textbf{R}  & 0.03           & 0.03          & 0.04          & 0.09          & \textbf{0.15} \\ \cline{2-7} 
\multicolumn{1}{|c|}{}                                                                                 & \textbf{F1} & 0.05           & 0.04          & 0.07          & \textbf{0.10}  & 0.11          \\ \hline
\multicolumn{1}{|c|}{\multirow{4}{*}{\textbf{SampleHST}}}                                                & \textbf{J}  & 0.40           & 0.59           & 0.72          & 0.75          & 0.88          \\ \cline{2-7} 
\multicolumn{1}{|c|}{}                                                                                 & \textbf{P}  & \textbf{0.84}  & \textbf{0.83} & \textbf{0.86} & 0.92          & 0.80          \\ \cline{2-7} 
\multicolumn{1}{|c|}{}                                                                                 & \textbf{R}  & 0.10           & 0.18          & 0.37          & 0.91          & \textbf{0.94} \\ \cline{2-7} 
\multicolumn{1}{|c|}{}                                                                                 & \textbf{F1} & 0.17           & 0.30          & 0.52          & \textbf{0.92} & 0.87          \\ \hline
\end{tabular}
\end{table}

We now focus on the case with high budget (10\%). Firstly, we see that SampleHST easily outperforms the PERCH-based method considering the primary evaluation criteria recall. Secondly, when we consider representative sampling, we see that the Jain score produced by SampleHST is {1.6}$\times$ better than the PERCH-based method. The reason for SampleHST performing better is as follows. The primary objective of SampleHST is to sample as much as anomalous traces possible. In high budget cases, it only shifts focus towards normal traces when the primary objective is fulfilled. Anomalous traces can create many groups, each with a small size, whereas normal traces create a small number of large groups. This is indeed the case with the production data. As a result, when SampleHST samples most of the traces from anomalous groups, it satisfies the demands of majority of the groups, making it more fair which is reflected in the Jain score.

\begin{figure}[t]
    \centering
     \subfloat[][$h_1=h_2$]{ \includegraphics[width=0.5\linewidth]{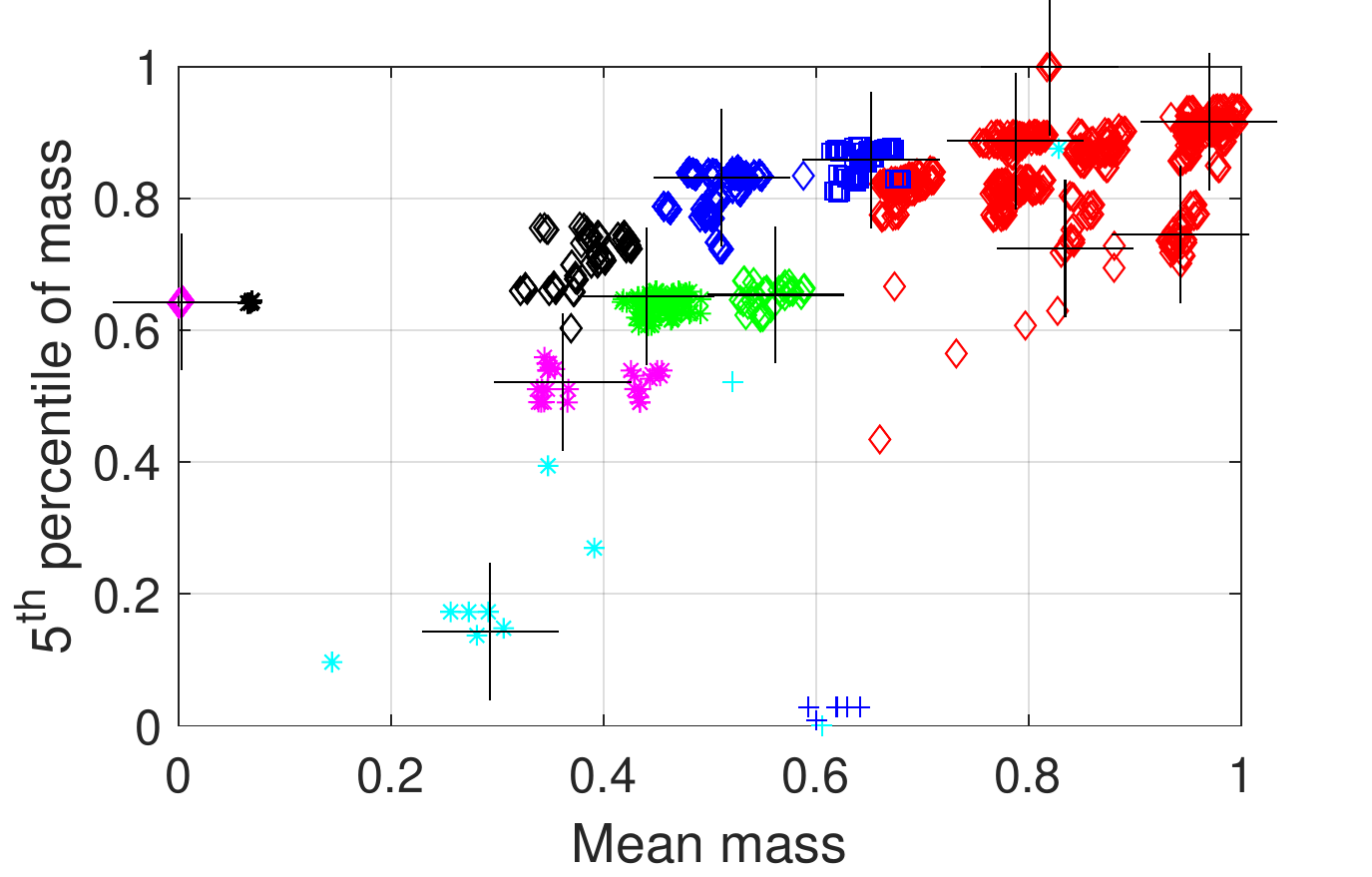}\label{fig:clust_hyper_cubes}}
     \subfloat[][$h_1 \neq h_2$]{\includegraphics[width=0.5\linewidth]{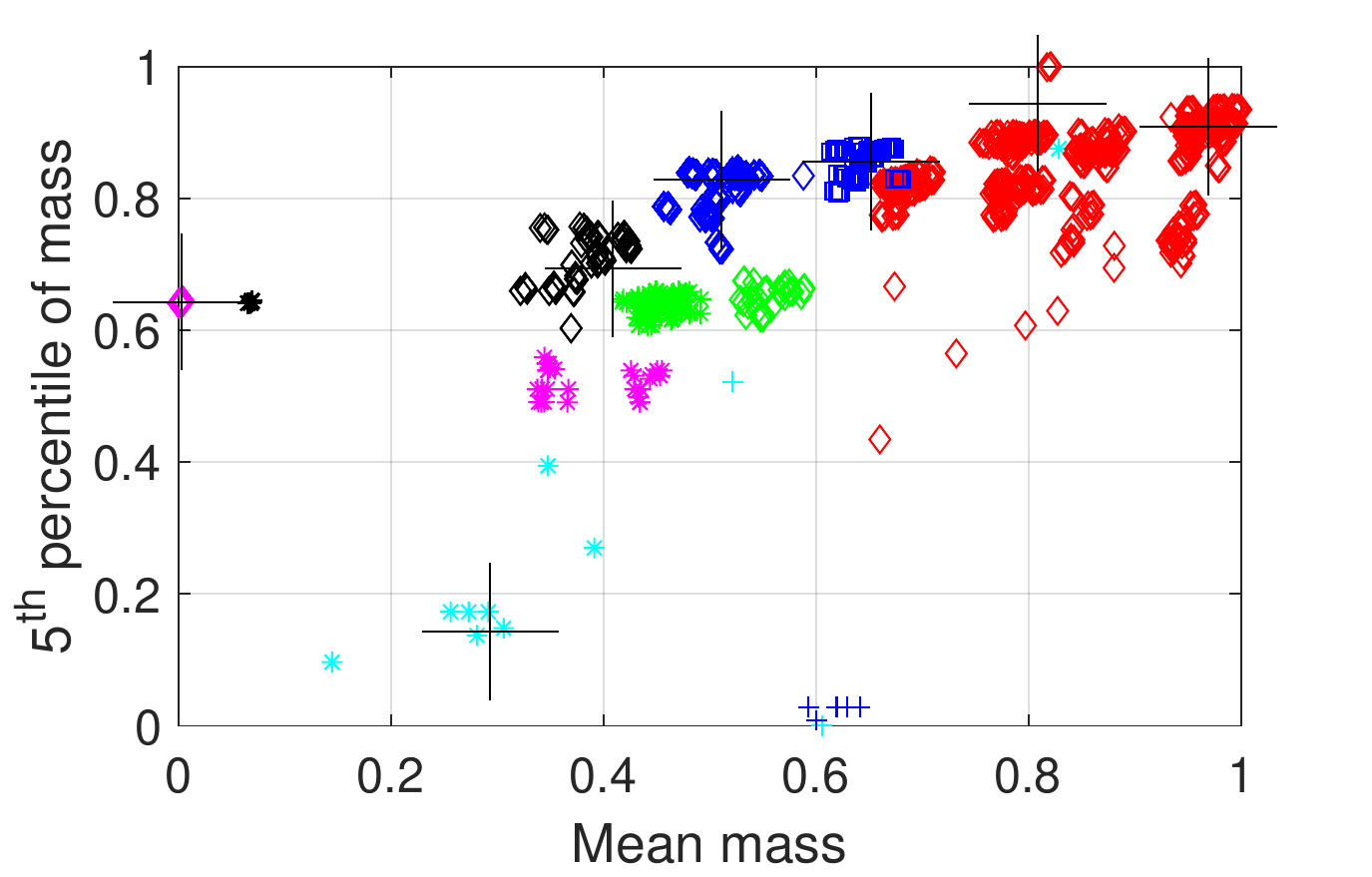}\label{fig:clust_hyper_rectangles}}
    \caption{Comparing clusters with equal and unequal clustering bandwidth}
    \label{fig:cube_vs_rectangle}
\end{figure}

\begin{table}[t]
\footnotesize
\centering
\caption{Performance of SampleHST  considering hyper-rectangles}
\label{tbl:results-sensitivity-march02}
\begin{tabular}{|r|c|c|c|c|}
\hline
\multicolumn{1}{|c|}{\textbf{}} & \textbf{Jain} & \textbf{Precision} & \textbf{Recall} & \textbf{F1-Score} \\ \hline
\textbf{0.05, 0.1}	&	0.76	&	0.90	&	0.91	&	0.91	\\ \hline
\textbf{0.05, 0.2}	&	0.75	&	0.91	&	0.91	&	0.91	\\ \hline
\textbf{0.05, 0.3}	&	0.74	&	0.93	&	0.91	&	0.92	\\ \hline
\textbf{0.1, 0.2}	&	0.74	&	0.94	&	0.91	&	0.92	\\ \hline
\textbf{0.1, 0.3}	&	0.73	&	0.97	&	0.92	&	0.95	\\ \hline
\end{tabular}
\end{table}
\noindent {\bf SampleHST with Hyper-Rectangles.} 
The mass scores work as anomaly signals to the SampleHST, which are not always likely to be equally strong in all clustering dimensions. In such cases, the traces may not be segregated ideally in that dimension. This is not a problem as long as we can separate anomalous traces from normal ones. However, if the bandwidth in that dimension is small, we can have multiple clusters in a particular region in the clustering hyper-plane, which represents traces of similar types. Thus rather than using a small clustering bandwidth in that dimension, as illustrated in Fig.~\ref{fig:cube_vs_rectangle}, we can chose a large one to remove clusters containing similar traces, allowing a more precise clustering. In other words, we can opt for hyper-rectangles, with unequal clustering bandwidths in each dimension, instead of hyper-cubes. When we observe the clustering status, as presented in Fig. \ref{fig:cube_vs_rectangle}, indeed with hyper-rectangles there are less number of clusters in the top right corner, that represents normal traces. Having less number of traces reduces the probability of sampling from normal groups, which is essential in low and moderate budget cases. This is also reflected in the sampling performance. In Table~\ref{tbl:results-sensitivity-march02} we present the results, for the 5\% budget case and for different sizes of hyper-rectangles. From these results, we can appreciate that the F1-Score for bandwidth $[0.1, 0.3]$ reaches $0.95$, which is higher than the one we achieved for hyper-cubes presented in Table \ref{tbl:results-compare-march02}. Moreover, and considering the hyper-rectangle $[0.1, 0.3]$ as our baseline we can see in Table~\ref{tbl:results-sampling-production-shapes} that the hyper-rectangles approach yields significantly better results in the metrics considered. In particular for low-budget scenarios we achieve on average an improvement of $1.12\times$ with respect to hyper-cubes.

\begin{table}[t]
\footnotesize
\centering
\caption{Sampling Results With Hyper-Cubes and Hyper-Rectangles}
\label{tbl:results-sampling-production-shapes}
\begin{tabularx}{\columnwidth}{|X|X|X|X|X|X|X|X|X|}
\hline
\multirow{2}{*}{} & \multicolumn{4}{c|}{\textbf{$h=0.1$}} & \multicolumn{4}{c|}{\textbf{$[h_1,h_2]=[0.1,0.3]$}} \\ \cline{2-9} 
 & \multicolumn{1}{c|}{\textbf{J}} & \multicolumn{1}{c|}{\textbf{P}} & \multicolumn{1}{c|}{\textbf{R}} & \textbf{F1} & \multicolumn{1}{c|}{\textbf{J}} & \multicolumn{1}{c|}{\textbf{P}} & \multicolumn{1}{c|}{\textbf{R}} & \textbf{F1} \\ \hline
\textbf{0.5\%} & 0.40 & \textbf{0.84} & 0.10 & 0.17 & 0.41 & \textbf{0.94} & 0.10 & 0.18 \\ \hline
\textbf{1\%} & 0.59 & \textbf{0.83} & 0.18 & 0.30 & 0.50 & \textbf{0.95} & 0.21 & 0.34 \\ \hline
\textbf{2\%} & 0.72 & \textbf{0.86} & 0.37 & 0.52 & 0.47 & \textbf{0.96} & 0.41 & 0.58 \\ \hline
\textbf{5\%} & 0.75 & 0.92 & 0.91 & \textbf{0.92} & 0.73 & 0.97 & 0.92 & \textbf{0.95} \\ \hline
\textbf{10\%} & 0.88 & 0.80 & \textbf{0.94} & 0.87 & 0.88 & 0.79 & \textbf{0.94} & 0.86 \\ \hline
\end{tabularx}
\end{table}

\section{Conclusion and Future Work}
\label{sec:conc_fairhst}

In this paper we propose a novel sampling method for distributed tracing namely \textit{SampleHST}. The objective of SampleHST is to take its sampling decision based on the proportion of sampling budget and the fraction of expected anomalous traces. If the budget is lower, the priority is to sample the anomalous traces. On the other hand, when the budget higher, the normal traces are sampled as well. This sampling process is based on an online clustering mechanism. The traces are first clustered using their mass scores generated using a forest of HST. After that, if the budget permits, the sampling decisions are taken based on the association of a trace with a cluster, where the clusters more likely to contain anomalous traces are prioritized. Our experiments, that considers production data from a cloud data center, show that SampleHST by far outperforms the recent approach targeting point anomalies.

A possible line of future research direction could be integrating the continuous trace properties, like the response time, to identify also the latency anomalies in an integrated approach.

\section*{Acknowledgments}
A. Gias is a commonwealth scholar, funded by the UK government. This research has received funding by Huawei Technologies (Ireland) Co., Ltd.


\appendix

\begin{proof}

Assume that we have $N$ observations $\{x_i\}_{i=1}^N$, with each data point having $d$ dimensions, that is $x_i=(x_{i,1},...,x_{i,d})$, then we can define the density gradient estimate $\hat{\nabla}p(x) \equiv \nabla\hat{p}(x)$ at $x=(x_1,...,x_d)$ as
\begin{equation}
\label{eq:additive_kernel_est}
    \begin{aligned}
        \nabla\hat{p}(x)
        =\frac{1}{Nh^d}\sum_{i=1}^N\nabla K \left(\frac{x_1-x_{i,1}}{h}, \ldots, \frac{x_d-x_{i,d}}{h}\right)\\
        =\frac{1}{Nh^d}\sum_{i=1}^N\sum_{k=1}^d\frac{\partial}{\partial x_k}K \left(\frac{x_1-x_{i,1}}{h}, \ldots, \frac{x_d-x_{i,d}}{h}\right)\cdot e_k
    \end{aligned}
\end{equation}
where $\hat{p}(x)$ is the kernel density estimator \cite{baruah2012evolving, langrene2019fast} for an unknown density $p$, $h$ is the bandwidth in all $d$ dimension, $K(\cdot)$ is the kernel function and $e_k$ is the $k$-th standard unit vector. Now using \eqref{eq:additive_kernel} as our kernel function, we get

\begin{equation}
\label{eq:diff_kernel}
    \begin{aligned}
        \sum_{k=1}^d\frac{\partial}{\partial x_k}K \left(\frac{x_1-x_{i,1}}{h}, \ldots, \frac{x_d-x_{i,d}}{h}\right)\cdot e_k\\
        =\sum_{k=1}^d\frac{\partial}{\partial x_k} \left [\frac{3}{d2^{d+1}}\sum\limits_{l=1}^d \left( 1-\left(\frac{x_l-x_{i,l}}{h}\right)^2\right)\right]\cdot e_k\\
        = \frac{3}{d2^{d+1}}\sum\limits_{k=1}^d \frac{\partial}{\partial x_k} \left( 1-\left(\frac{x_k-x_{i,k}}{h}\right)^2\right)\cdot e_k\\
        = \frac{3}{d2^{d}h^2}\sum\limits_{k=1}^d  \left(x_{i,k}-x_k\right)\cdot e_k
    \end{aligned}
\end{equation}

Substituting $\nabla K(\cdot)$ in \eqref{eq:additive_kernel_est} with \eqref{eq:diff_kernel}, the density gradient estimate is
\begin{equation}
\label{eq:density_gradient}
    \begin{aligned}
        \nabla\hat{p}(x) =\frac{1}{Nh^d}\frac{3}{d2^{d}h^2}\sum_{i=1}^N \sum\limits_{k=1}^d  \left(x_{i,k}-x_k\right)\cdot e_k\\
    = \frac{N_x}{Nh^d}\frac{3}{d2^{d}h^2}\sum_{k=1}^d \frac{1}{N_x} \sum\limits_{x_{i,k} \in S_r(x)}  \left(x_{i,k}-x_k\right)\cdot e_k\\
    = \frac{N_x}{Nh^d}\frac{3}{d2^{d}h^2}\sum_{k=1}^d \left(\mu_k-x_k\right)\cdot e_k\\
    \end{aligned}
\end{equation}

Here, in \eqref{eq:density_gradient}, $N_x$ is the number of data points for a region $A_h(x)$, $\mu_k$ is the mean of all the data points in that region along the $k^{\text{th}}$ dimension. Note that, the volume of the region $A_h(x)$ is $h^d$ and the probability density estimate $\hat{p}(x)$ over the region using a uniform kernel is $\frac{N_x}{Nh^d}$.

The objective of the mean shift algorithm is to move away from the valley and towards the function mode. This can be achieved through gradient ascent. That is, for two consecutive iteration $t$ and $t+1$ the shift in the variable $x_j$ in $k^{\text{th}}$ dimension can be expressed as 

\begin{equation}
\label{eq:grad_ascent}
x_{j,k}^{t+1} = x_{j,k}^{t}+c\frac{\nabla\hat{p}(x)}{\hat{p}(x)}
\end{equation}

Now substituting the value of $\hat{p}(x)$ and $\nabla\hat{p}(x)$ and considering $c = \frac{d2^{d}h^2}{3}$ in \eqref{eq:grad_ascent}, we show in \eqref{eq:shift} that the shift is equal to the mean of $x$ in dimension $k$. This mean that, from the online clustering perspective, with the arrival of each sample, the cluster center shifts $u_k$ in dimension $k$.

\begin{equation}
\label{eq:shift}
     x_{j,k}^{t+1} 
    = x_{j,k}^{t}+\left(\mu_k -x_{j,k}^t\right)\\
    = \mu_k
\end{equation}

Based on \eqref{eq:shift}, we conclude that the mean shift algorithm is applicable to the kernel in \eqref{eq:additive_kernel} when the cluster bandwidth $h$ is equal in all dimensions.

\end{proof}

\end{document}